\documentclass[12pt]{article}
\usepackage{epsf}
\usepackage{graphicx}
\usepackage{lscape}
\usepackage{amssymb}
\usepackage{pazh}
\usepackage{natbib}

\tightenlines

\begin{document}

{\footnotesize Astronomy Letters, Vol. 33, No. 5, 2007, pp. 299-308.
Translated from Pis'ma v Astronomicheskii Zhurnal, Vol. 33, No. 5,
2007, pp. 340-351.}

\title{\bf High-Mass X-ray Binaries and the Spiral Structure of the Host Galaxy}

\author{P. E. Shtykovskiy$^{1,2 \,*}$, M. R. Gilfanov$^{2,1}$}

\affil{
$^{1}$Space Research Institute, Profsoyuznaya str. 84/32, Moscow 117997, Russia\\
$^{2}$MPI for Astrophysik, Karl-Schwarzschild str. 1, Garching, 85741, Germany}

\sloppypar
\vspace{2mm}
\noindent

We investigate the manifestation of the spiral structure in the distribution
 of high-mass X-ray
binaries (HMXBs) over the host galaxy. We construct the simplest kinematic model. 
It shows that the
HMXBs should be displaced relative to the spiral structure observed in such traditional
 star formation
rate indicators as the H$_{\alpha}$ and far-infrared emissions because of their finite 
lifetimes. Using Chandra
observations of M51, we have studied the distribution of X-ray sources relative
 to the spiral arms of this
galaxy observed in H$_{\alpha}$. Based on K-band data and background source number counts,
 we have separated
the contributions from high-mass and low-mass X-ray binaries and active galactic nuclei. 
In agreement
with model predictions, the distribution of HMXBs is wider than that of bright HII regions 
concentrated
in the region of ongoing star formation. However, the statistical significance of this result 
is low, as is
the significance of the concentration of the total population of X-ray sources to the spiral 
arms. We also
predict the distribution of HMXBs in our Galaxy in Galactic longitude. The distribution 
depends on the
mean HMXB age and can differ significantly from the distributions of such young objects as 
ultracompact
HII regions.


\vfill

{$^{*}$ E-mail: pavel@hea.iki.rssi.ru}
\newpage
\thispagestyle{empty}
\setcounter{page}{1}

\section*{INTRODUCTION}

Chandra and XMM-Newton X-ray observations
of nearby galaxies have revealed rich populations
of compact X-ray sources in them (see, e.g., Fabbiano
2006; Kilgard et al. 2005). Their spectral analysis,
flux variability, luminosity functions, and spatial
distribution make it possible to identify them with
known (in our Galaxy) classes of objects -- high-mass
and low-mass X-ray binaries, supernova remnants,
ultrasoft X-ray sources, and to distinguish a hitherto
unknown class of ultraluminous X-ray sources.

From the standpoint of their relation to star formation,
the X-ray sources can be divided into two
groups: young objects, such as high-mass X-ray binaries
(HMXBs) and supernova remnants (SNRs),
and old objects, such as low-mass X-ray binaries
(LMXBs). It would be natural to expect the former
to concentrate in regions of recent star formation and
the latter to follow the galactic stellar mass distribution.
Indeed, Chandra observations of nearby galaxies
show that the populations of X-ray sources in starforming
galaxies differ radically from those in elliptical
galaxies. This is particularly clearly seen when the
luminosity functions of these sources are analyzed:
the luminosity functions for the former are in the form
of a power law with an index of 1.6, as for the HMXBs
in our Galaxy, while those for the latter are similar
in form to the luminosity function of LMXBs in our
Galaxy (Grimm et al. 2003; Gilfanov 2004).

In nearby spiral galaxies, two components should
be observed in the distribution of X-ray sources:
HMXBs concentrating to the spiral arms and LMXBs
distributed more smoothly with the maximum density
at the galactic center. On the one hand, Chandra
observations of such galaxies as M51 and M101
suggest that the X-ray sources actually concentrate
to the spiral arms. However, a more detailed analysis
indicates that, in reality, the picture is more complex.
First, HMXBs reflect the star formation in a galaxy
that took place $\sim5-60$~Myr ago, i.e., strictly speaking,
they are not an instantaneous star formation rate
(SFR) indicator (Shtykovskiy and Gilfanov 2005).
Such times may turn out to be important from the
standpoint of galactic dynamics, since the characteristic
revolution time of the stars in a typical spiral
galaxy is $\sim200$~Myr. As a result, the spiral structure
observed in the HMXB distribution will be distorted
and displaced relative to that observed, for example,
in H$_{\alpha}$. On the other hand, the density contrast
between the arms and the interarm space in the stellar
population of certain galaxies can reach $\sim3$. This will
give rise to a spiral structure in the old population of
LMXBs as well.

In this paper, we construct a kinematic model for
the spatial distribution of X-ray binaries in a spiral
galaxy, compare its predictions with Chandra observations
of M51, and make predictions for our Galaxy.

\section{THE SPIRAL STRUCTURE IN VARIOUS SFR INDICATORS}
\label{sec:spiralind}

The basic principles of the modern theory of spiral
structure in galaxies were laid down by Lin and Shu
(1964, 1966) and Lin et al. (1969), who suggested
the hypothesis of a quasi-stationary density wave.
According to this hypothesis, the spiral structure of a
galaxy is a manifestation of the density wave - a stellar
disk and gas density perturbation that propagates
through the galaxy and that does not decay for a long
period. Although the reasons why a stationary density
wave emerges and is maintained are unknown, there
are many candidates for their role, such as the influence
of a neighboring galaxy, asymmetry at the galactic
center, etc. Since the velocity dispersion in the
interstellar gas is comparatively low, the amplitude of
the density wave triggered by a gravitational potential
perturbation can be large. In contrast, the amplitude
of the density wave in the stellar disk should be small
(at least for an isolated galaxy). Moreover, a shock
wave leading to a very narrow zone of gas compression
with a large density jump can generally emerge in
the gas (Roberts 1969). The gas compression triggers
star formation in it; as a result, young stars producing
a distinct galactic spiral structure in the optical band
are formed.

In addition to the visible range, the spiral structure
can also be observed in many other SFR, stellar
mass, and gas indicators. The H$_{\alpha}$, ultraviolet, and
far-infrared emissions produced by massive stars, the
near-infrared emission from old stars, the 21-cm HI
emission, and the CO emission are the most important
ones. Simple considerations suggest that the
spiral structure will be different in different indicators.
For example, the H$_{\alpha}$ emission originates in HII regions
containing young stars with masses 10~M$_{\odot}$,
while the ultraviolet emission is associated with the
photospheric emission of stars from a wide mass
range (Kennicutt 1998). Since the angular velocities
of the stellar disk and the density wave are different,
young stars are displaced from the density wave shock
front with time. Due to short lifetimes of the massive
stars responsible for the H$_{\alpha}$ emission, 20 Myr, its
peak should lie not far from the spiral density wave
front. On the other hand, the less massive stars responsible
for the ultraviolet emission have lifetimes
up to 100 Myr and can be displaced significantly from
their birthplaces. As a result, the spiral structure in
the ultraviolet will be considerably wider than that
in H$_{\alpha}$. Displacements of this kind between various indicators
were actually observed, for example, between
H$_{\alpha}$ and the ultraviolet in M51 (Petit et al. 1996), HI
and the nonthermal radio continuum inM51 and M83
(Tilanus and Allen 1991), etc.

The situation with HMXBs is similar to that
described above. Indeed, the lifetime of an optical
counterpart with a mass of  $\sim$8~M$_{\odot}$ is $\sim40$~Myr,
i.e., HMXBs reflect the star formation that took
place several tens of Myr ago. A clear example of
this, though unrelated to the spiral structure, is the
absence of correlation between the surface density of
HMXBs and the H$\alpha$ emission intensity in the Large
Magellanic Cloud (Shtykovskiy and Gilfanov 2005).

\subsection{The Spiral Structure in the Distribution of High-Mass X-ray Binaries}
\label{sec:hmxbspiral}

To summarize the aforesaid, we conclude that
the following factors affect the spatial distribution of
HMXBs.

(i) The spatial distribution of star-forming regions.
The theory of density waves and observations of spiral
galaxies show that a logarithmic spiral can serve as
a good approximation for the location of the spiral
density wave front. The location of a spiral arm is then
specified in polar coordinates  (r, $\Theta$) by the relation
\begin{equation}
\Theta-\Theta_0=\ln(r/r_0)/\tan\psi,  
\label{eq:eq2}
\end{equation}
where $\psi$ is the angle between the spiral and the tangent
to the circumference called the pitch angle and
(r$_0,\Theta_0)$ is a point on the spiral, for example, its origin.
We assume that the star formation takes place in a
narrow region along this spiral.

(ii) The dynamics of HMXBs. The HMXBs born
at the shock front are drawn into the overall motion
of the stellar disk in accordance with the galactic
rotation curve. To a first approximation, it can be
represented as the motion in circular orbits around
the galactic center. The spiral density wave rotates
with the pattern speed $\Omega_p$ in the same direction as the
stellar disk. The HMXB locations relative to the instantaneous
location of the spiral density wave front,
which is initially equal to zero, in time $\tau$ is then given
by the following equation:
\begin{equation}
\Delta\Theta=(\Omega(r)-\Omega_p)\tau,  
\label{eq:eq1}
\end{equation}
where $\Delta\Theta$ is the displacement of the objects relative
to the instantaneous location of the density
wave and $\Omega(r)$ is the galactic rotation curve. In other
words, this formula gives the locations of objects
with age $\tau$ relative to the currently observed starforming
region. Obviously, the HMXBs move faster
than the density wave within the corotation radius 
r$_{cr}$ ($\Omega(r_{cr})=\Omega_p$) and lag behind it outside r$_{cr}$.

In this formula, the gravitational field perturbation
in the galaxy and the fact that the stars move
in epicyclic orbits, not in circular ones, are ignored.
To accurately describe the dynamics of the various
components of a spiral galaxy, we must also consider
the dynamics of gas clouds and the star formation in
them under the effect of gravitational instability. Models
of this interaction (see, e.g., Leisawitz and Bash
1982; Roberts and Stewart 1987) yield contradictory
results: young stars will move almost along the spiral
arm for a certain time in some of the models and will
be rapidly displaced from it in other models.

(iii) The evolution of the HMXB number with time
elapsed since the star formation event. A detailed
investigation of this question is a difficult task that
requires developing population synthesis models and
is beyond the scope of this paper. Here, we only
briefly note the main factors that specify the evolution
time scale of the population of HMXBs. First, this
is the time it takes for a black hole or a neutron
star to be formed. Here, two natural time scales can
be identified: the lifetime of the most massive star
($\approx$100~M$_{\odot}$) corresponding to the formation of the first
compact object (black hole), t$_{min}\approx2-3$~Myr, and the
lifetime of the least massive star  ($\approx$8~M$_{\odot}$) capable of
forming a compact object -- the formation time of the
last neutron star,  t$_{max}\approx40$~Myr. The evolution of the
companion star until the onset of an active phase,
mass accretion onto the compact object, is of no less
importance. The duration of this phase can also be
rather long, up to several tens of Myr, depending
on the mass of the companion star. In addition, the
duration of the X-ray activity phase, which should be
much shorter than the first two phases, $10^{3}-10^{6}$~yr,
must be taken into account. As a result, the maximum
age for a HMXB can reach more than 40 Myr.
This is also confirmed by observations: for example,
analysis of the population of HMXBs in the Magellanic
Clouds showed that binaries with ages up to
$\sim$60~Myr could be present among them (Shtykovskiy
and Gilfanov 2007). Below, as the time scale for the
existence of a HMXB population, we use both a conservative
estimate of t$_{max}\approx40$~Myr and larger values,
demonstrating the magnitude of the effect in the latter
case for the oldest binaries.

\begin{figure*}
\begin{center}
\includegraphics[width=0.8\textwidth,clip=true]{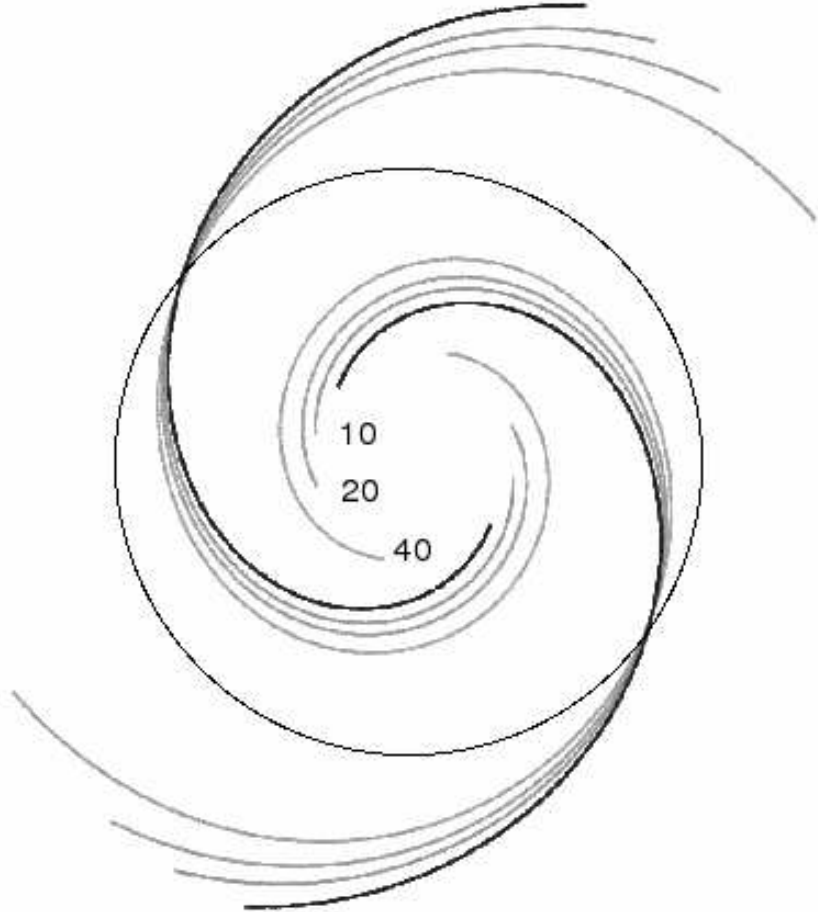}
\end{center}
\caption{Spiral structure in the distribution of HMXBs.
Different curves corresponds to the locations of objects
with ages of 10, 20, and 40 Myr calculated using Eq. (2).
The location of the region of ongoing star formation
(black curve) is also shown. The circumference indicates
the location of the corotation radius.}
\label{fig:fig1}
\end{figure*}

It is also worth noting that the displacement of
HMXBs relative to other SFR indicators observed in
galaxies at distances $\ga1~$~Mpc can be somewhat smaller
than that in nearby galaxies. Indeed, when X-ray
binaries are observed in distant galaxies, there is a
selection effect: the brightest objects are selected.
Since the luminosity of a HMXB is related to the
mass of its optical counterpart (Postnov 2003), a
mass selection effect also arises, i.e., predominantly
the highest-mass and, accordingly, youngest stars
for which the displacement from the spiral arms is
smaller, will be observed. However, apart from the
mass of the counterpart, the luminosity also depends
on the orbital size. This reduces the selection effect
and increases the contribution from ``old'' HMXBs to
the observed population of X-ray sources. A detailed
analysis of this effect is beyond the scope of this paper.

To estimate the displacement of X-ray sources
relative to the galactic spiral structure, we calculated
the expected spatial distributions of HMXBs with
ages of 10, 20, and 40 Myr. As the region of ongoing
star formation, we used the approximation of the inner
part of the star-forming region in M51 (see below)
by a logarithmic spiral ($\psi=19~\deg$). The displacement
of HMXBs relative to the instantaneous location of
the spiral was calculated using Eq. (2) by assuming
that the rotation curve and the density wave pattern
speed also corresponded to those of M51. The derived
distributions together with the current location of the
spiral structure are shown in Fig. 1.

\begin{figure*}
\hbox{
\includegraphics[width=0.5\textwidth]{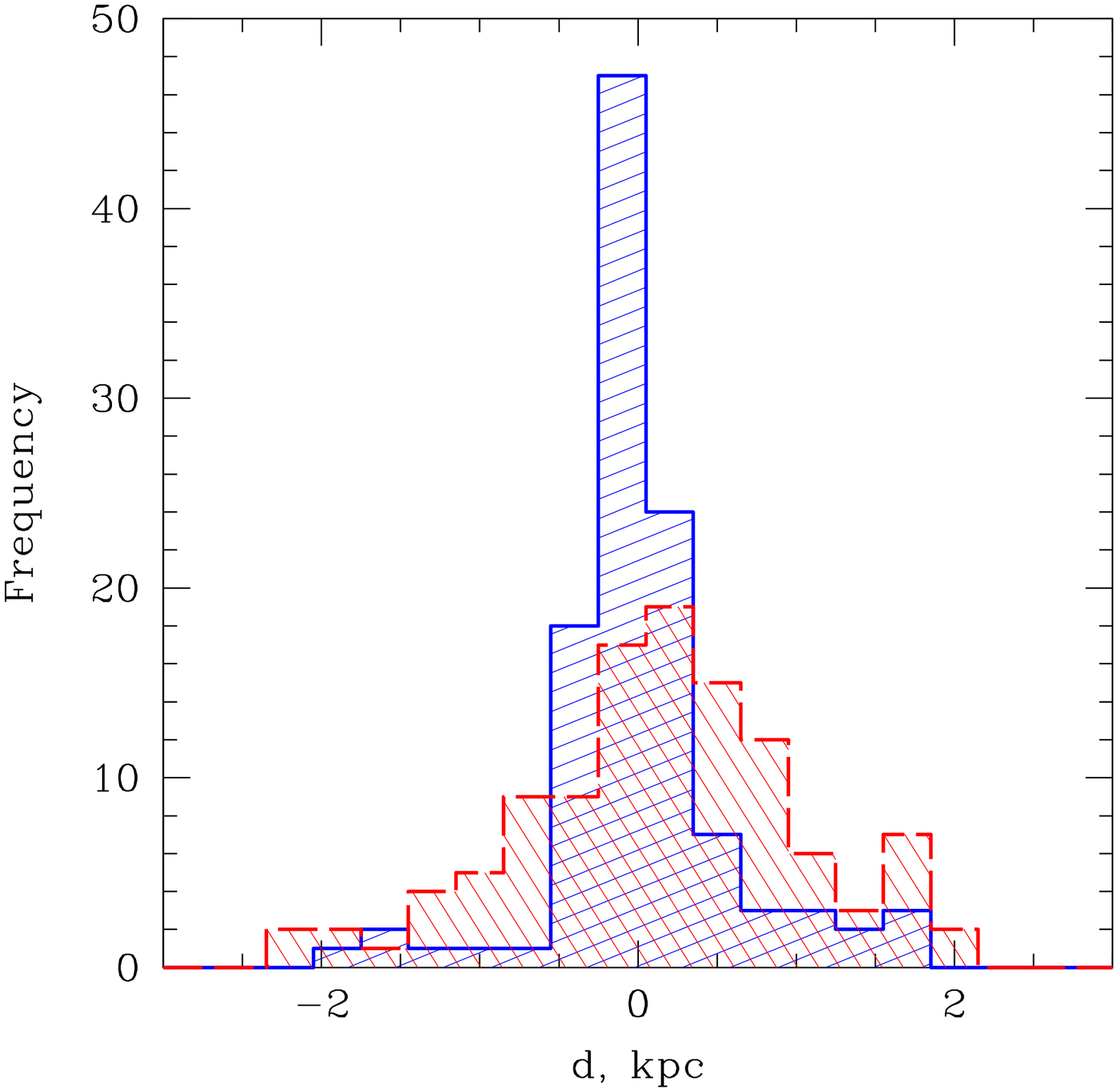}
\includegraphics[width=0.5\textwidth]{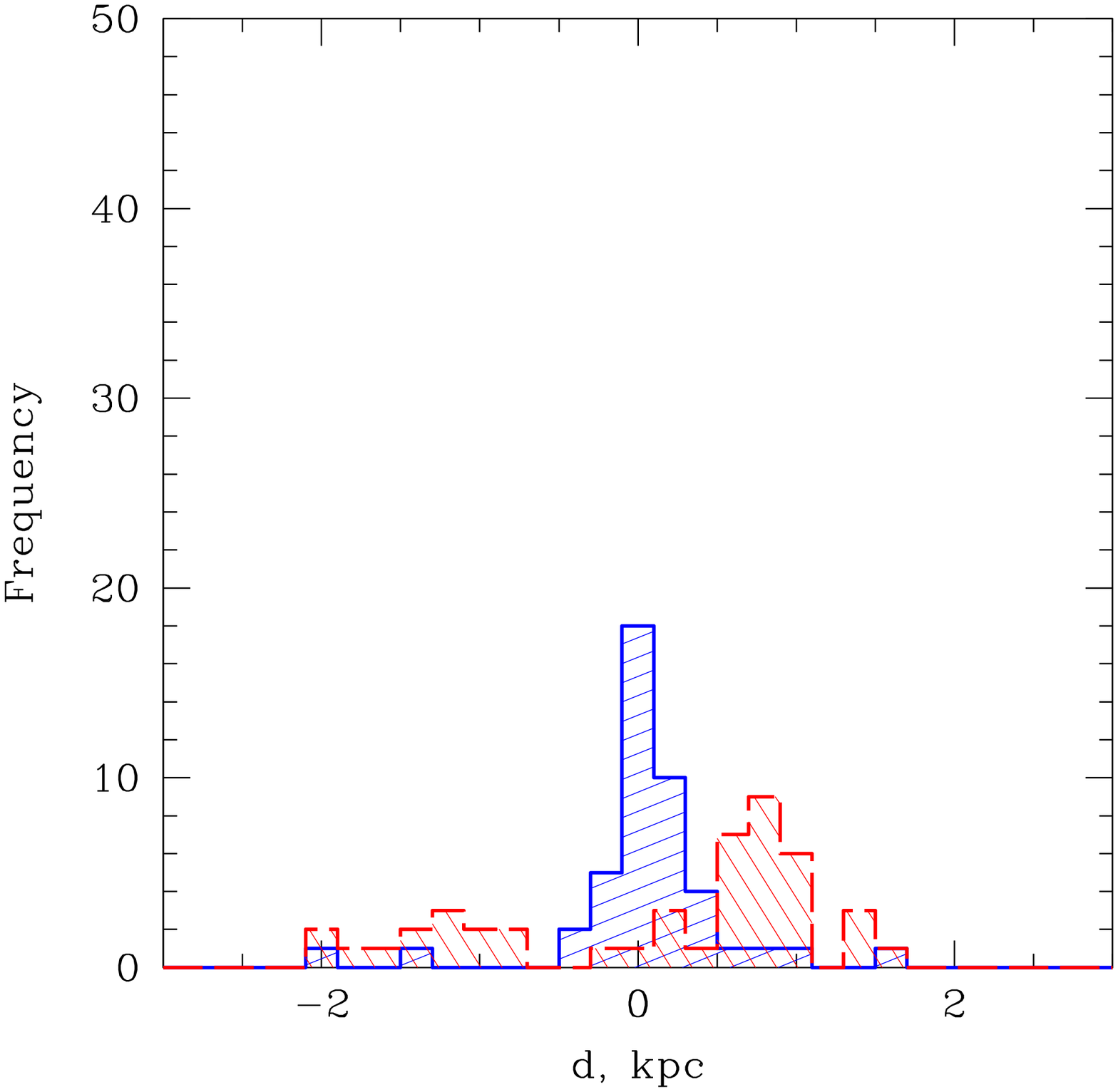}
}
\caption{(a) Distribution of distances from bright HII regions 
(Petit et al. 1996) to the nearest spiral arm (solid histogram)
in M51. Also shown is the distribution for objects with an age of 
40 Myr (dotted histogram). (b) The same for objects within
the corotation radius.}
\label{fig:fig3}
\end{figure*}

\section{THE SPIRAL STRUCTURE AND HIGH-MASS X-RAY BINARIES: COMPARISON WITH OBSERVATIONS}
\label{sec:observations}

Various methods can be used to study the spiral
structure in the observed distribution of sources
or emission intensity at a certain wavelength. For
example, the distribution of displacements of the
polar angle relative to the spiral angle, $\chi=\phi-\ln(r/r_0)/\tan\psi$, 
can be constructed. In this presentation
method, a logarithmic spiral with two arms
is projected into two $\delta$-functions near 0 and $\pi$.
However, when the shape of the spiral structure is
far from a logarithmic spiral, it is convenient to use
the distributions of source distances to it. Obviously,
this distribution will change in pattern as the sources
``grow old''. This effect can be demonstrated with an
actual example using bright HII regions, for which
the distribution of distances to the nearest spiral arm
characterizes the locations of the youngest objects.
The distribution for older objects can be obtained
by displacing their locations relative to the current
location of the spiral in accordance with Eq. (2).
For this kind of demonstration, we chose bright
HII regions in M51 ((S$>1.5\times10^{-17}$~W/m$^2$) from
the catalog by Petit et al. (1996) and displaced them
by assuming that the spiral was a trailing one (i.e.,
the rotation is counterclockwise; see Fig. 3). We
then constructed the distributions of distances from
the original and displaced sources to the nearest
spiral arm (for more detail on the spiral structure
in M51, see below) for sources in a wide range of
galactocentric distances, 2.3~kpc$<$r$<$9.5~kpc, and
within the corotation radius, 2.3~kpc$<$r$<$5.6~kpc.
A distortion of the distribution is clearly seen in both
cases (see Fig. 2); in the latter case, it is in the form
of a displacement.

Obviously, for comparison with the qualitative
model constructed above, a galaxy with active star
formation and a low stellar mass whose spiral structure
is close to a logarithmic spiral with a large
pitch angle would be ideal. In addition, it is desirable
that the angular size of the galaxy should be small
enough for the contribution from background sources
to be at a minimum.

\subsection{Comparison with observations: M51}
\label{sec:m51}

One of the most suitable galaxies for our purpose
among those observed by Chandra is M51. This
galaxy has a distinct spiral structure and is located at
a distance of 9.7~Mpc (Sandage and Tammann 1975).
The displacements of various indicators of the spiral
structure in M51 clearly indicate that we are dealing
with a density wave (see, e.g., Tilanus and Allen 1989,
1991). To determine the orientation of M51 in space,
we take the coordinates of its center RA(J2000)=$13^h29^m52^s.71$ and 
DEC(J2000)=$47\deg25\arcmin42\arcsec.6$ (Ford et al.
1985), P.A. = 170$\deg$, and its inclination to the plane
of the sky i=20$\deg$ (Tully 1974). As the spiral wave
pattern speed, we use $\Omega_p=38$~km/s/kpc obtained
by Zimmer et al. (2004) by the Weinberg-Tremain method. 
Data on the rotation curve were
taken from Tilanus and Allen (1991) and fitted by the
law $V(r)\propto\sqrt{(r/r_0)^{1.3}/(1+(r/r_0)^{2.3})}$. The rotation
curve and the density wave pattern speed correspond
to a corotation radius r$_{cr}\approx5.6$~kpc, in agreement
with the values obtained by other methods (Vogel et al. 1993).

An important peculiarity of M51 is its interaction
with its companion, NGC 5195. As a result, it is often
cited as an example of a galaxy where the spiral structure
can be excited and maintained by tidal forces
(see, e.g., Toomre 1978). The tidal interaction is probably
also responsible for the unusually high density
contrast in the K band and, hence, in the stellar
population of M51 (Rix and Rieke 1993). Obviously,
a high density contrast in the stellar disk can change
radically the manifestation of the spiral structure in
X-ray binaries. For example, it would be natural to
expect the appearance of a spiral structure in the
distribution of not only HMXBs, but also LMXBs.

The spiral structure in M51 is more complex in
shape than a simple logarithmic spiral and can be
approximated by the latter only in a limited range of
radii. Moreover, the ionized gas velocity field on the
periphery of M51 indicates that the outermost parts
of the spiral rotate with a pattern speed different from
that of the inner spiral (Vogel et al. 1993).

Below, we exclude the region around NGC 5195
with a radius of $\approx1.7\arcmin$ from our analysis.

\subsubsection{X-ray sources toward M51.}
\label{sec:m51xraysrc} 

For our analysis,
we used the list of sources detected by Chandra
(Terashima and Wilson 2004). The total number
of sources within a galactocentric radius of
$\approx10$~kpc is 88 at the limiting sensitivity 
S$_{lim}=1.4\times10^{-15}$~erg/cm$^2$/s
in the 0.5--8~keV energy
band (for a photon index of 1.5 and N$_{H}=1.3\times10^{20}$~cm$^{-2}$), 
corresponding to the luminosity 
L$_X=1.6\times10^{37}$~erg/s. As with any other galaxy, most
of the X-ray sources toward M51 belong to one of
the following classes: HMXBs, LMXBs, SNRs, and
background active galactic nuclei (AGNs). Below,
we discuss the contribution from each of them to the
population of X-ray sources toward M51.

(1) AGNs. According to the background source
number counts by Moretti et al. (2003), N$^{S}_{CXB}\approx$13
cosmic X-ray background (CXB) sources are expected
in a circle with a radius of 3.4$\arcmin$ if we use the
source counts in the soft 0.5--2~keV energy channel
(S$^{S}_{lim}=4.62\times10^{-16}$~erg/cm$^2$/s, assuming
$\alpha=1.5$ and N$_{H}=1.3\times10^{20}$cm$^{-2}$) and 
N$^{H}_{CXB}\approx$22 
background sources are expected if we use the source
counts in the hard 2--10~keV energy channel (S$^{H}_{lim}=1.17\times10^{-15}$~erg/cm$^2$/s) 
with fluxes above the detection threshold.

A similar value, N$_{CXB}\approx$18, is obtained if we use
the source counts in the 0.5--8~keV energy band
obtained by Kim et al. (2006). We use this number
below.

(2) HMXBs. To estimate the expected number
of HMXBs, we use a calibration from Grimm
et al. (2003) with a modified normalization (Shtykovskiy
and Gilfanov 2005):
\begin{equation}
N(>L)=1.8\times SFR(L_{38}^{-0.6}-210^{-0.6}),
\label{eq:eq6}
\end{equation}
where SFR is the star formation rate of the host
galaxy.

According to the IRAS catalog (Rice et al. 1988),
the far-infrared luminosity of L$_{FIR}$=$1.8\times10^{10}$L$_{\odot}$.
The L$_{FIR}$-SFR calibration (Kennicutt 1998)
yields SFR$\approx3.1$~M$_{\odot}$/yr. Using Eq. (3), we then
obtain the expected number of HMXBs with luminosities
above the detection threshold S$_{lim}$(2--10)$=1.17\times10^{-15}$~erg/cm$^2$/s,
N$_{HMXB}\approx19$. A slightly
larger number is obtained if we use the star formation
rate derived from the ultraviolet luminosity
of M51, SFR(UV)$\approx4.3$~M$_{\odot}$/yr (Calzetti et al.
2005), which corresponds to N$_{HMXB}\approx27$ HMXBs.
However, this number is less reliable, because the
absorption in the ultraviolet is uncertain.

It is worth noting that the calibration by Grimm
et al. (2003) also includes the possible contribution
from SNRs and ultrabright sources, which should be
similar to HMXBs in spatial distribution.

(3) LMXBs. To estimate the number of LMXBs
in M51, we use the fact that their number is proportional
to the stellar mass of the galaxy (Gilfanov
2004). The latter, in turn, is proportional to
the K-band flux (Bell and de Jong 2001). Using
K-band images of M51 from the 2MASS Large
Galaxy Atlas (Jarrett et al. 2003) and assuming that
the magnitude in a 2$\arcmin$ aperture is K$(<2\arcmin)=6.76$
(Rix and Rieke 1993), we obtain K(r$<9.5$~kpc)=6.4 
for the entire galaxy. According to the calibration
by Bell and de Jong (2001) and the color
(B--V)$_0$=0.56 (de Vaucouleurs et al. 1991), this
corresponds to M$_{*}=2.63\times10^{10}$M$_{\odot}$. Using the
galactic stellar mass -- LMXB number calibration
(Gilfanov 2004), we then obtain  N$_{LMXB}\approx$33 (22 in
the ring 2.3~kpc$<$r$<$9.5~kpc, 13 within the corotation
radius 2.3~kpc$<$r$<$5.6~kpc) LMXBs with
luminosities above the threshold value, 
S$_{lim}$(2--10)$=1.17\times10^{-15}$~erg/s/cm$^2$.

Our estimates roughly agree with the observed
total number of sources (see the table). The distribution
of X-ray sources in M51 is shown in Fig. 3. It
should be kept in mind that since the nucleus has a
high X-ray brightness, Terashima and Wilson (2004)
excluded the central part of the galaxy from their analysis.
Therefore, the central depression in the number
of sources should not be interpreted as the absence of
X-ray binaries in this region.

\begin{table}
\caption{X-ray sources toward M51}
\renewcommand{\arraystretch}{1.2}
\label{tb:src}
\begin{centering}
\begin{tabular}{lcccc}
\hline
Source class & HMXB & LMXB & CXB \\
\hline
Expected number of sources & 19-27 & 33 & 13-22 \\
\hline
Total observed number & & & \\
of sources & & 88 & \\
\hline
\end{tabular}\\
\end{centering}
\smallskip
\end{table}

\begin{figure*}
\begin{center}
\includegraphics[width=0.8\textwidth,clip=true]{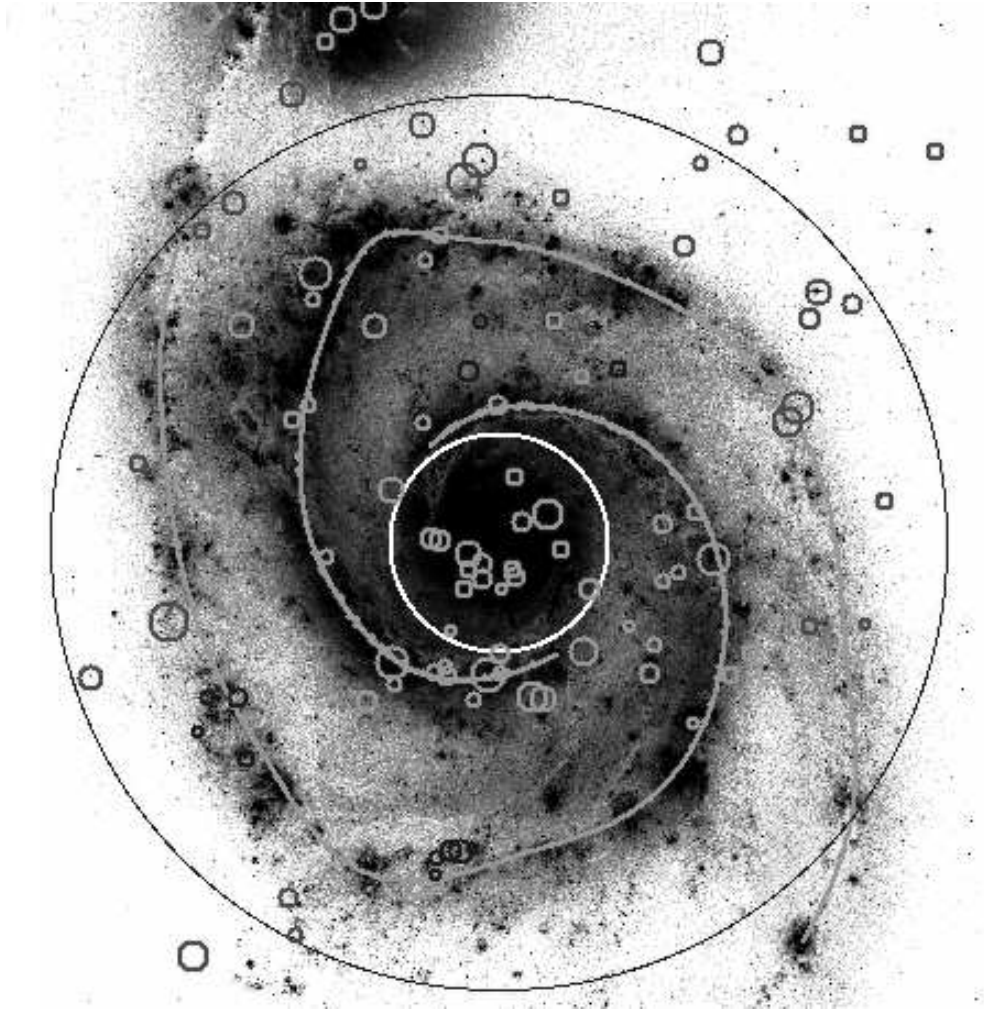}
\end{center}
\caption{HST H$_{\alpha}$ image of M51 (Mutchler et al. 2005).
The circles mark the positions of the X-ray sources detected
by Chandra (Terashima and Wilson 2004); the
circle radius reflects the source X-ray luminosity. For a
higher contrast, the circles have different colors in different
regions. Also shown are the two arms of the spiral
structure used as the location of the region of ongoing
star formation. The two circles around the galactic center
(the light and dark ones at small and large distances,
respectively) indicate the region from which the sources
are taken for our analysis (2.3-9.5~kpc).}
\label{fig:fig2}
\end{figure*}

The spiral structure in the distribution of
X-ray binaries. It is natural to investigate the spiral
structure in the distribution of X-ray binaries with
respect to a certain indicator of the region of ongoing
star formation or, in other words, an indicator of the
shock front location. For example, the location of
the most massive (and, accordingly, youngest) stars,
the inner edge of the dust lane clearly seen in the
B band, the gas velocity distribution, etc. can serve
as such an indicator. In our case, it is convenient to
use the H$_{\alpha}$ emission peak, which essentially reflects
the location of the youngest stars, as the location of
the spiral. For this purpose, we used the HST mosaic
of M51 (Mutchler et al. 2005). The locations of the
spiral arms drawn along the H$_{\alpha}$ emission peak 
are indicated in Fig. 3 by the solid line. In addition, we
used bright HII regions 
(S$>1.5\times10^{-17}$W/m$^2$)
from the catalog by Petit et al. (1996) as an additional
indicator. 

To estimate the degree of concentration of Xray
sources to the spiral arms, we calculated the
distribution of their distances to the nearest spiral
arm using sources at galactocentric distances 2.3--9.5~kpc. 
Positive and negative distances are assigned
to sources above and below the spiral, respectively.
The derived distribution (corrected for the galaxy inclination)
is shown in Fig. 4.

Despite the obvious physical meaning of the constructed
distribution, the interpretation of its shape
is nontrivial. Indeed, a solid angle can be associated
with each interval of distances on it. As the solid angles
for different intervals can differ, the distributions
in this representation will be distorted, more specifically,
they will be more concentrated to zero than
they actually are. For example, a uniform distribution
of sources, such as AGNs, will have a maximum
near zero whose distinctness will be determined by
the shape of the spiral and by the boundaries of the
region from which the X-ray sources are taken. As
a result, the observed distribution will have a clear
meaning only when it is considered in comparison
with another, ``calibration'', distribution. Obviously,
in our case, the distribution of bright HII regions and
a uniform distribution are such distributions.

\begin{figure*}
\hbox{
\includegraphics[width=0.5\textwidth]{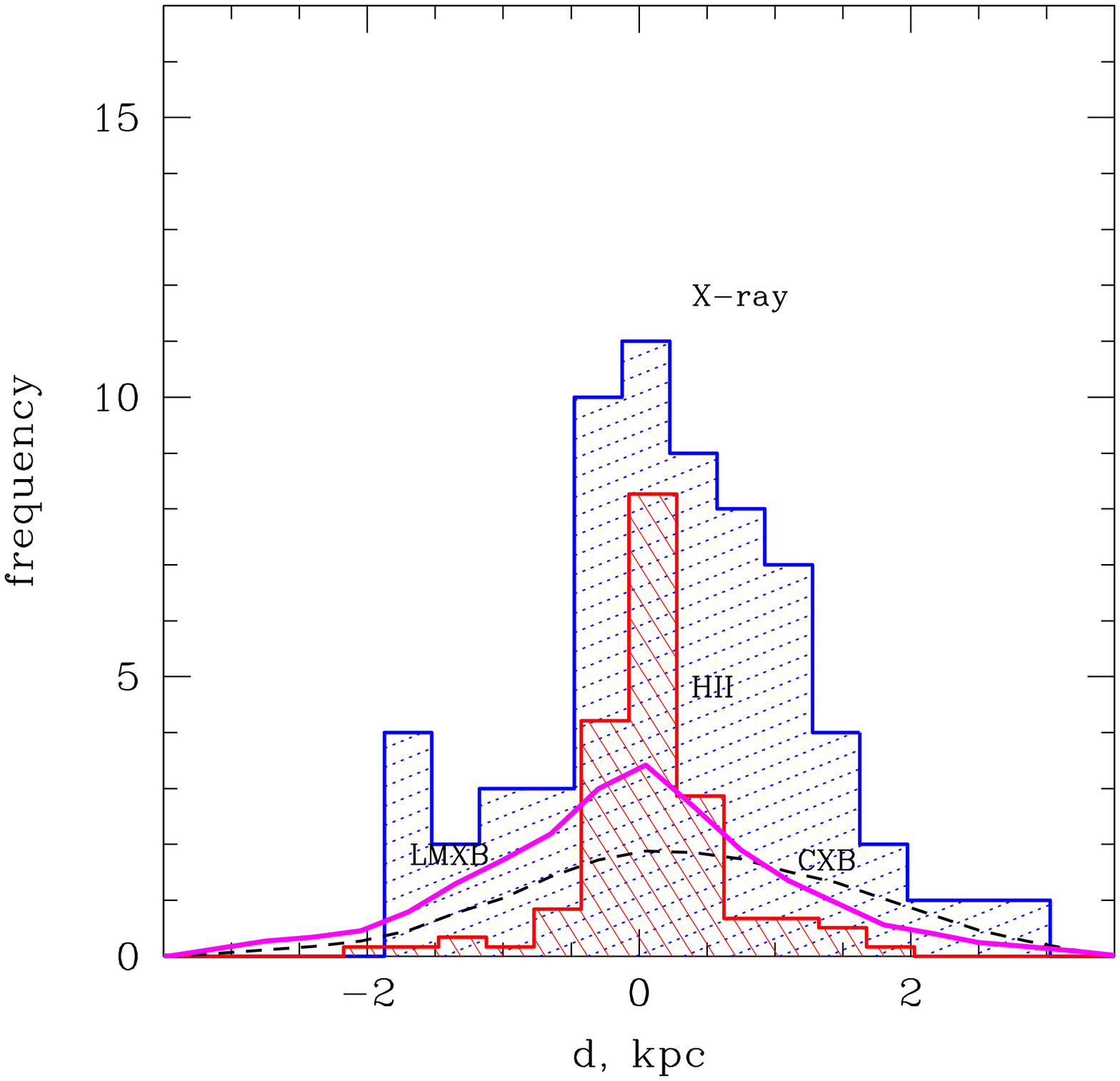}
\includegraphics[width=0.5\textwidth]{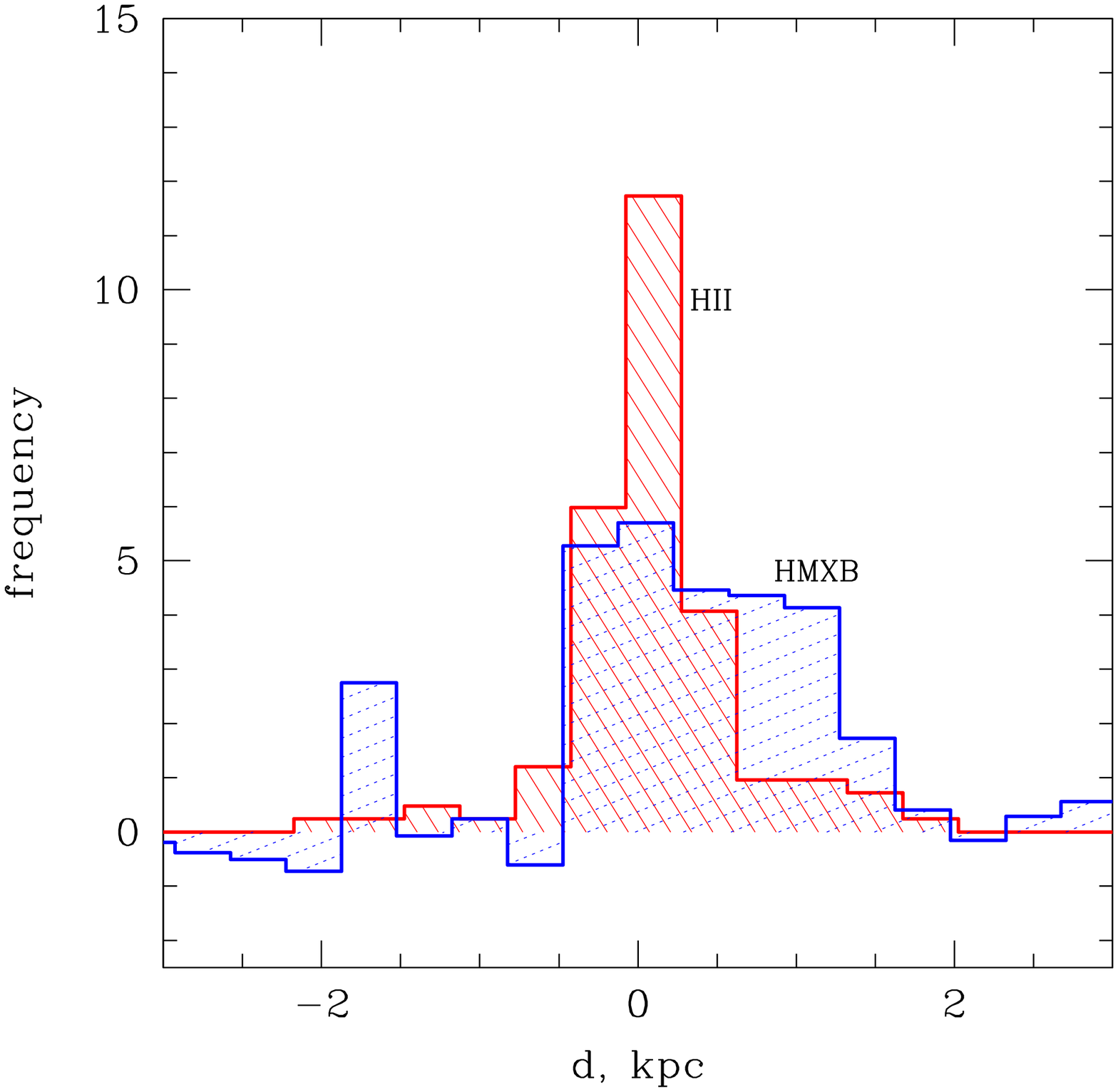}
}
\caption{(a)
Distribution of distances to the nearest spiral arm for X-ray sources 
(wide histogram) and bright HII regions (narrow
histogram, reverse hatching). 
The distribution for HII regions is normalized to the difference 
between the total number of X-ray
sources and the expected number of LMXBs and AGNs. 
The solid and dotted curves correspond to the predictions for LMXBs
and AGNs, respectively. (b) The difference between the distribution of X-ray sources  
and the total distribution of LMXBs and
AGNs (wide histogram) and the distribution of bright HII regions (narrow histogram, 
reverse hatching).
}
\label{fig:fig5}
\end{figure*}

The observed distribution is determined by the
following factors.

(1) Finite width of the star-forming region. Even
if the star formation in gas clouds is assumed to be
triggered at a narrow shock front, the star formation
process is not an instantaneous event, which smears
the spatial region of zero-age stars. As was noted
above, we consider our distribution relative to the distribution
of HII regions, which, obviously, will reflect
the width of the star-forming region. To make a quantitative
comparison of the distribution for HMXBs
with the distribution for HII regions possible, we normalized
the latter to the difference between the total
number of X-ray sources and the expected number of
LMXBs and AGNs.

(2) Dynamics of HMXBs. The distortion of the
distance distribution due to the displacement of
sources relative to the region of ongoing star formation
was demonstrated above. The expected distribution
of HMXBs with allowance made for their
dynamics is shown in Fig. 2. A characteristic feature
of the distribution is its asymmetry, which is
particularly noticeable if only the sources within the
corotation radius are taken into account. In addition,
the distribution of HMXBs is spread due to the initial
velocities acquired through supernova explosions.

(3) LMXBs. As was noted above, a high stellar
density contrast in M51 ($\delta\sigma/\sigma\sim$2--3, mainly within
the corotation radius) should lead to a concentration
of these sources to the spiral arms. Obviously,
their spatial distribution should differ from the observed
one for HMXBs. To estimate the distribution
of LMXB distances to the nearest spiral arm, we
generated a model population of sources distributed
over the galaxy in accordance with the K-band flux.
The number of sources is taken to be large ($\approx10^5$)
to suppress the contribution from the Poisson noise.
The derived distribution for the model sources was
normalized in accordance with the expected number
of LMXBs,  N$_{LMXB}=22$.

(4) Contribution from background sources. To estimate
the contribution from background sources, we
calculate the distribution of distances to the nearest
spiral arm for sources distributed uniformly over the
sky and normalize it to the predictions of the source
counts (see above), N$_{CXB}$=18.

The above components are shown in Fig. 4. Obviously,
the distribution of X-ray sources is wider and
more asymmetric than that of HII regions. This is
the result of a cumulative effect from LMXBs, AGNs,
and HMXB dynamics. To estimate the contribution
from HMXBs, we give the difference between the
distribution of X-ray sources and the sum of the
predicted distributions for LMXBs and AGNs. The
derived distribution is also asymmetric; it is similar
in pattern to the prediction of the simplest model for
the displacement of HMXBs (Fig. 2). However, the
result is statistically insignificant. The significance of
the fact that the distribution of X-ray sources is not a
sample from the total distribution of LMXBs, AGNs,
and bright HII regions calculated by Kuiper's test is
less than 2$\sigma$ (p$\approx13\%$).

Interestingly, the formal significance of the concentration
of X-ray sources to the spiral arms is also
low, although the reverse seems to be true at first
glance at Fig. 3. Indeed, judging by Fig. 3, the density
of sources near the spiral arms clearly exceeds their
density in the interarm space. If we do a quantitative
count, then we will find that almost as many sources
are observed within a strip of $\pm$600~pc as outside it,
34 versus 32. At the same time, the corresponding
areas are 11.1 versus 19.7 square arcminutes. Thus,
the concentration is actually present, but its statistical
significance is low. The significance of the fact
that the distribution of distances of X-ray sources
to the nearest spiral arm is not a sample from the
distribution of sources distributed uniformly over the
sky calculated by Kuiper's test is also found to be
low, about  2$\sigma$ for sources with 2.3~kpc$<$r$<$9.5~kpc
and slightly higher for sources within the corotation
radius. Thus, the existing number of sources is clearly
not enough to reach firmconclusions about the distribution
of HMXBs relative to the spiral arms in M51.
Moreover, as we said above, M51 is one of the best
examples for studying the spiral structure among the
galaxies observed by Chandra. Although a comprehensive
analysis of the archival data on other spiral
galaxies is beyond the scope of this paper, we expect
the statistical significance of the manifestation of a
spiral structure in X-ray sources in them to be even
lower. Further deep Chandra observations of spiral
galaxies with intense star formation are required to
investigate the manifestation of the spiral structure in
the distribution of X-ray sources in more detail.

\begin{figure*}
\begin{center}
\includegraphics[width=0.8\textwidth,clip=true]{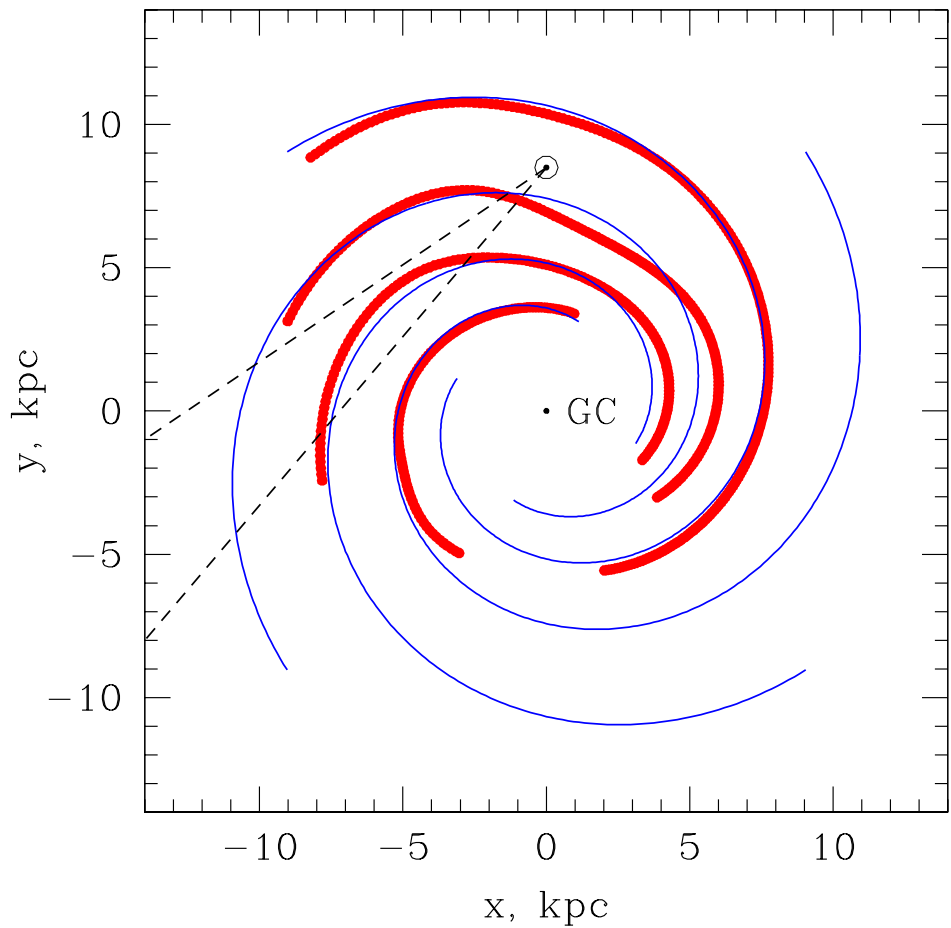}
\end{center}
\caption{Approximation of the spiral structure of our
Galaxy by a logarithmic spiral and a model of th spiral
structure from Taylor and Cordes (1993) (thick curve).
The location of the Solar system and the solid angle containing
the tangent to one of the spiral arms are shown.}
\label{fig:fig7}
\end{figure*}

\subsection{High-Mass X-ray Binaries in Our Galaxy}
\label{sec:galaxy}

The simplest kinematic model constructed here
also allows us to make predictions about the expected
distribution of HMXBs in our Galaxy observed from
the Solar system (r=8.5~kpc). For this purpose, we
assume that the spiral structure of our Galaxy follows
a logarithmic spiral with four arms and a pitch angle
$\psi=13\deg$ that originates at a Galactocentric distance
of 3.3 kpc (see, e.g., Vallee 2005, 1995) (Fig. 5).
Figure 5 also presents a model of the spiral structure
in the distribution of free electrons in the Galaxy
from Taylor and Cordes (1993). Obviously, its shape
deviates from a logarithmic spiral only slightly. We
assume that the SFR and, accordingly, the number of
HMXBs along the spiral arm length element decrease
with decreasing exponential Galactic disk density.
The characteristic width of the spiral arm is taken to
be w=0.3~kpc (FWHM=0.7~kpc). The SFR at an
arbitrary point in the Galaxy is calculated as follows:
\begin{equation}
SFR\propto \int e^{-r/r_0}\times e^{-d^2/2w^2}\times dl,
\label{eq:eqsfr}
\end{equation}
where the integral is taken along the spiral arms,
dl is the length element of the spiral arm, r is its
Galactocentric distance, d is the distance from the
point to the spiral arm, and r$_0$=3.9~kpc (Benjamin
et al. 2005).

After their birth, the HMXBs are displaced relative
to the instantaneous locations of the spiral arms in
accordance with Eq. (2). The rotation curve is assumed
to be flat in the galactocentric distance range
of interest, 
V/(220 km/s)=$a_1(R/8.5$~kpc$)^{a_2}+a_3$~ ($a_1=1.00767$, $a_2=0.0394$, $a_3=0.00712$)
 (Brand and Blitz 1993); as the spiral density wave pattern
speed, we take $\Omega_p$=24~km/s/kpc (Dias and
Lepine 2005), which corresponds to the corotation
radius r$_{cr}\approx9.3$~kpc.

Having integrated the SFR along the line of sights
originating from the Solar system, we obtained the
distribution of HMXBs in Galactic longitude. The
distributions for objects with ages of 0, 40, and 80~ Myr
constructed by assuming that we are capable of detecting
only sources at heliocentric distances smaller
than 6.5 and 11.5 kpc are shown in Fig. 6. The
detection limits roughly correspond to the depths of
the INTEGRAL survey for sources with luminosities
 L$_X=10^{35}$~erg/s and $10^{35.5}$~erg/s (Lutovinov et al. 2005).
As might be expected, the positions of the maxima in
the distribution for the youngest objects correspond
to the spiral arm tangents (see Fig. 5), since in this
case the integrated SFR along the line of sight is at a
maximum. At the same time, as we see from Fig. 5,
the maxima in the distribution of older objects can
be displaced significantly. Besides, additional peaks
related to the fact that the hitherto invisible inner
parts of the spiral arms become visible appear.

Obviously, the weakness of the constructed model
is that we do not know the exact shape of the spiral
arms and the SFR behavior along them. Indeed, if the
SFR is highly nonuniform along the spiral arms, then
the maxima in the distributions of young sources can
also be observed in directions different from the spiral
arm tangents.

The displacement of the HMXB maxima relative
to the locations of the spiral arm tangents were actually
observed by Lutovinov et al. (2005) (see Fig. 7
in this paper, the peak in a direction l$\approx+40\deg$). The
observed position of the peak corresponds to binaries
with ages $\approx40-80$~Myr, which corresponds to the
oldest binaries from the standpoint of HMXB evolution.
However, to reach specific conclusions, we
must primarily have a larger number of sources and,
in addition, understand what contribution the SFR
nonuniformity along the spiral arms makes to the
distribution.
hensive analysis of the archival data on other spiral
galaxies is beyond the scope of this paper, we expect
the statistical significance of the manifestation of a
spiral structure in X-ray sources in them to be even
lower. Further deep Chandra observations of spiral
galaxies with intense star formation are required to
investigate the manifestation of the spiral structure in
the distribution of X-ray sources in more detail.

\begin{figure*}
\begin{center}
\includegraphics[width=1.0\textwidth,clip=true]{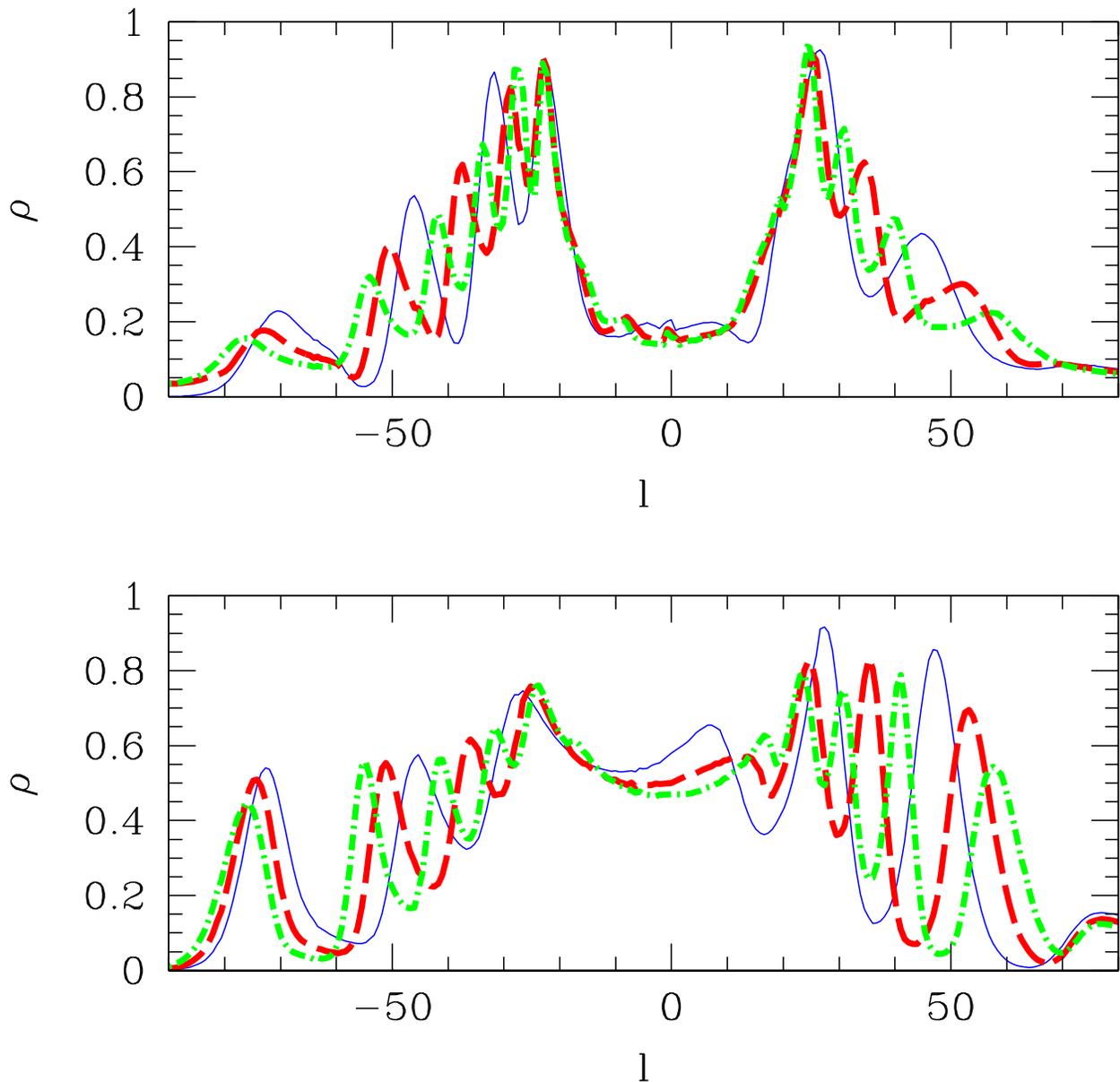}
\end{center}
\caption{Model distribution of HMXBs in our Galaxy in longitude for objects with ages of 0 (solid curve), 40 (dotted curve), and
80 Myr (dashed-dotted curve). Obviously, the solid curve describes the manifestation of the spiral structure in the youngest
indicators, such as the H$_{\alpha}$ line and the distribution of the most massive early-type stars. We see that the peaks in the
distribution of HMXBs are displaced considerably as the latter ``grow old''. 
The distributions were constructed for sources
closer than (a) 11.5 and (b) 6.5 kpc.}
\label{fig:fig8}
\end{figure*}

\section*{CONCLUSIONS}
\label{sec:summary}

We considered the spatial distribution of HMXBs
relative to the spiral structure of the host galaxy. We
constructed the simplest model for the kinematics of
HMXBs. It shows that they can be displaced to an
appreciable distance from the location of the shock
front at which ongoing star formation takes place over
their lifetimes (see Figs. 1 and 2). As a result, the
spiral structure in the distribution of HMXBs can be
displaced relative to the spiral structure observed in
classical SFR indicators, such as the H$_{\alpha}$ emission.
The displacement should be most pronounced for
binaries with neutron stars and minimal for binaries
with black holes. The evolution of theHMXB number
with time elapsed since the star formation event can
be judged by the displacement pattern.

As an illustration of this effect, we considered
the spatial distribution of X-ray sources in the
galaxy M51 by analyzing the distribution of their
distances to the spiral arms. We showed that the
components attributable to HMXBs and LMXBs
and background AGNs could be identified in the
distribution using K-band data and background
source number counts. The distribution of HMXBs
shows a clear tendency for them to concentrate to
the spiral arms. In agreement with predictions of the
kinematic model, it is wider and more asymmetric
than the distribution of bright HII regions, which
reflects the region of ongoing star formation (see
Figs. 2 and 4). However, the statistical significance of
both the concentration of X-ray sources to the spiral
arms and the fact that the distribution of HMXBs
differs from that of bright HII regions is low ($\la2\sigma$).
Further deep Chandra observations of spiral galaxies
are required to reach firm conclusions about the
distribution of HMXBs relative to the spiral arms.

We also calculated the expected distribution of
HMXBs in our Galaxy observed from the Solar
system. We showed that the maximum number of
sources could be observed in directions different from
the directions tangential to the spiral arms, where the
integrated current SFR is at a maximum (Fig. 6),
because the HMXBs are displaced relative to the
shock front. The displacement pattern allows the
peculiarities of the HMXB distribution observed by
Lutovinov et al. (2005) to be explained qualitatively.

\begin{acknowledgements}

P.E. Shtykovskiy thanks the European Association
for Research in Astronomy (EARA;
MEST-CT-2004-504604) for support by a Marie
Curie grant and the Max-Planck-Institut fu¨ r Astrophysik,
where this work was performed, for hospitality.
This work was also supported by grant no. 
NSh-1100.2006.2 from the President of Russia
and the Russian Academy of Sciences (``Origin and
Evolution of Stars and Galaxies'' Program).

\end{acknowledgements}

\pagebreak

\end{document}